\documentclass[twocolumn,showpacs,prb,superscriptaddress,floatfix]{revtex4}

\usepackage{color} 
\usepackage{tabularx} 
\usepackage{epsfig}
\usepackage{amsmath} 
\usepackage{amssymb} 
\usepackage{graphicx}
\usepackage{wasysym}
\usepackage{oldgerm}
\usepackage{psfrag}

\newcommand{\reduce}{.96}

\begin{document}

\title{Critical behavior and universality in L\'evy spin glasses}

\author{Juan Carlos ~Andresen}
\affiliation {Theoretische Physik, ETH Zurich, CH-8093 Zurich,
Switzerland}

\author{Katharina ~Janzen}
\affiliation {Institut f\"ur Physik, Carl-von-Ossietzky-Universit\"at,
26111 Oldenburg, Germany}

\author{Helmut G.~Katzgraber} 
\affiliation {Department of Physics and Astronomy, Texas A\&M University,
College Station, Texas 77843-4242, USA}
\affiliation {Theoretische Physik, ETH Zurich, CH-8093 Zurich, Switzerland}

\date{\today}

\begin{abstract}

Using large-scale Monte Carlo simulations that combine parallel
tempering with specialized cluster updates, we show that Ising
spin glasses with L\'evy-distributed interactions share the
same universality class as Ising spin glasses with Gaussian or
bimodal-distributed interactions. Corrections to scaling are large
for L\'evy spin glasses. In order to overcome these and show that
the critical exponents agree with the bimodal and Gaussian case, we perform an
extended scaling of the two-point finite size correlation length and
the spin-glass susceptibility.  Furthermore, we compute the critical
temperature and compare its dependence on the disorder distribution
width to recent analytical predictions [J.~Stat.~Mech.~(2008) P04006].

\end{abstract}

\maketitle

\section{Introduction}
\label{sec:intro}

Although universality has been established for many systems without
disorder and frustration, there are still skeptics that question this
cornerstone of the theory of statistical mechanics when applied to
disordered spin systems with frustrated interactions. According to
universality, the values of quantities such as critical exponents,
do not depend on microscopic details of the model, but only on
e.g., the space dimension and the symmetry of the order parameter.
Arguments based on high-temperature series expansions\cite{daboul:04}
support universality and there is no a~priori reason why systems with
both disorder and frustration, such as spin glasses,\cite{binder:86}
might not show universal features.  However, numerical studies are
difficult\cite{ogielski:85a, ogielski:85, mcmillan:85, singh:86,
bray:85, bhatt:85, bhatt:88, kawashima:96, bernardi:96, inigues:96,
berg:98, marinari:98, palassini:99b, mari:99, ballesteros:00,
mari:01, mari:02, nakamura:03, pleimling:05, joerg:06, ostilli:06,
campbell:06, toldin:06, katzgraber:06, hasenbusch:08b, roma:10}
and suffer from strong corrections to scaling. Therefore, there is
still debate\cite{bernardi:96,mari:99,mari:01,henkel:05,pleimling:05}
for some model systems if the shape of the disorder distribution can
influence the universality class of the system.

Although it is now well established that universality
is not violated for nearest-neighbor spin glasses
with compact disorder distributions (e.g., Gaussian or
bimodal),\cite{katzgraber:06,hasenbusch:08b,joerg:06} some studies
suggest that this might not be the case when the disorder distributions
are broad.\cite{mari:01} If the spin interactions are drawn from
a Gaussian or bimodal distribution the probability to have extremely
large interactions is very small. It is, however, unclear if strong
couplings between the spins change the universality class of the
system.  Selecting the interactions between the spins from a L\'evy
distribution allows one to continuously tune the probability to have
very strong bonds in the system.  In particular, for $\alpha < 2$
(see below for details) the L\'evy distribution has broad tails and
thus the probability to have a strong bond between two spins is large,
especially in the limit $\alpha \to 1$.

Using large-scale Monte Carlo simulations that combine parallel
tempering with specialized cluster moves,\cite{janzen:08} as well as
extended scaling techniques,\cite{campbell:06} our results show that
L\'evy spin glasses do obey universality for the system sizes studied.
Our estimates of the critical exponents agree within error bars
with the best known estimates\cite{katzgraber:06,hasenbusch:08b} for
Gaussian and bimodal disorder.  Furthermore, we probe recent analytical
predictions\cite{janzen:08} made for the critical temperature of
L\'evy spin glasses as a function of the disorder distribution width.

The paper is structured as follows: In Sec.~\ref{sec:model}
we introduce the model studied, as well as the measured
observables. Section \ref{sec:alg} outlines the special (cluster)
algorithm used to treat strong interactions in the L\'evy spin glass,
the finite-size scaling analysis, and how we estimate the critical
temperature, followed by results presented in Sec.~\ref{sec:results},
as well as concluding remarks.

\begin{figure}[!tbh]
\includegraphics[width=\reduce\columnwidth]{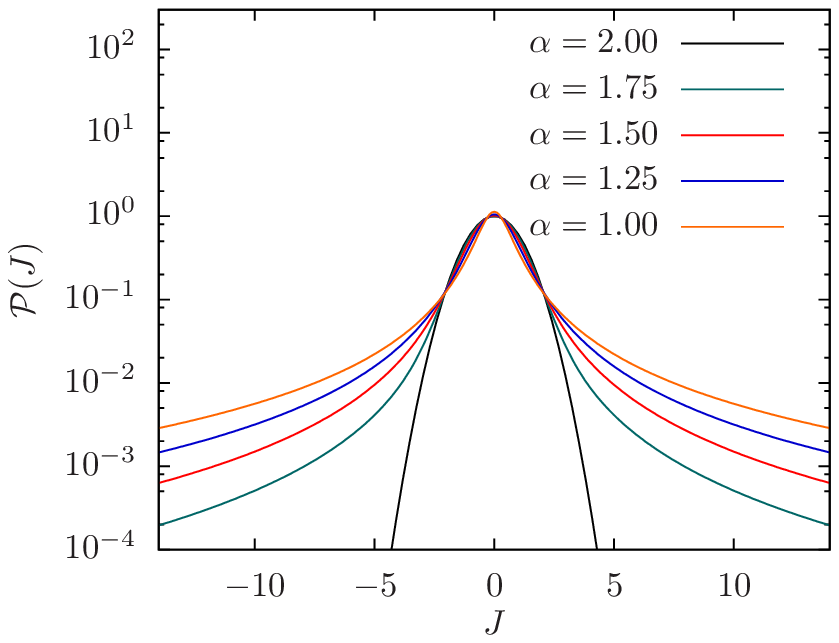}
\caption{(Color online) 
L\'evy distribution ${\mathcal P}(J)$ for the different values of
the shape parameter $\alpha$ and $c=1/\sqrt{2}$, as studied here. In
particular, for $\alpha = 2$ a Gaussian distribution is recovered. For
$1 \leqslant \alpha < 2 $ the distribution is fat-tailed, as can be
seen in the linear-log plot.
}
\label{fig:dist}
\end{figure}

\section{Model and Observables}
\label{sec:model}

We study the critical behavior of the Edwards-Anderson Ising spin
glass\cite{edwards:75} with L\'evy-distributed interactions, i.e.,
\begin{equation}
{\mathcal H} = - \sum_{\langle i,j\rangle} J_{ij}S_iS_j ,
\end{equation}
where the sites $i$ lie on a three-dimensional cubic lattice of
size $N=L^3$, $L$ the linear dimension, and the spins $S_i$ can take
the values $\pm 1$. Periodic boundary conditions are used to reduce
corrections to scaling. The sum is over nearest neighbors and the
interactions $J_{ij}$ are independent random variables taken from a
L\'evy distribution with zero mean and $c=1/\sqrt{2}$ defined through
the characteristic function $\phi(t)$ as
\begin{equation}
{\mathcal P}(J) =  
\frac{1}{2 \pi} \int_{-\infty}^{+\infty} dt \phi(t)e^{-it J} = 
\frac{1}{2 \pi} \int_{-\infty}^{+\infty} dt e^{-it J - |c t|^\alpha} . 
\label{eq:distribution}
\end{equation}
The parameter $\alpha$ influences the shape of the distribution
and, in particular, the width of the tails. For $\alpha = 2$
Eq.~(\ref{eq:distribution}) reduces to a Gaussian with variance
$\sigma = \sqrt{2} c$.  When $1 \leqslant \alpha < 2$ the tail of the
distribution decays as a power law, as seen in Fig.~\ref{fig:dist}.
In this case, exchange interactions with very large values can occur,
albeit with small probability. However, such strong interactions form
``dimers'' of spins that cannot be flipped with standard Monte Carlo
methods.\cite{hartmann:01} For decreasing $\alpha$ the probability to
have dimers grows, as well as the average of the maximum exchange
interaction value.

To test universality, at least two independent critical exponents
need to be computed.  Therefore, in the simulations, we measure the
following quantities.

The spin overlap $q$ is defined as
\begin{equation}
q = \frac{1}{N}\sum_i S_i^{(1)} S_i^{(2)} ,
\end{equation}
where $(1)$ and $(2)$ are two copies of the system with identical
disorder.  Using $q$ we define the Binder ratio\cite{binder:81} $g$ via
\begin{equation}
g = \frac{1}{2} 
\left(
3 - \frac{\left[\langle q^4\rangle\right]_{\rm av}}
{\left[\langle q^2\rangle\right]_{\rm av}^2}
\right) 
\sim {\tilde G}[L^{1/\nu}(T-T_c)] ,
\label{eq:binder}
\end{equation}
where $\langle \cdots \rangle$ represents a thermal average and
$[\cdots]_{\rm av}$ an average over the disorder. The Binder ratio is
a dimensionless function $\tilde{G}$, i.e., data for different system
sizes cross at a putative transition temperature $T_c$.  A finite-size
scaling analysis of the universal function\cite{privman:84} $\tilde{G}$
allows one to determine the critical exponent $\nu$ for the
correlation length.

The spin-glass susceptibility $\chi_{\rm SG}$ is defined as
\begin{equation}
\chi_{\rm SG} =	N \left[\langle q^2 \rangle\right]_{\rm av} \sim
L^{2-\eta}\tilde C [L^{1/\nu}(T-T_c)] .
\label{eq:susceptibility}
\end{equation}
A finite-size scaling analysis of the susceptibility thus permits
the calculation of the critical exponent $\eta$. However, a simple
scaling analysis of the spin-glass susceptibility suffers from strong
corrections to scaling\cite{katzgraber:06} and therefore an extended
scaling\cite{campbell:06} is performed below where the scaling
function incorporates corrections derived from the resummation of a
high-temperature series expansion.

Finally, we measure the two-point finite-size correlation
length.\cite{palassini:99b,ballesteros:00} To do so we introduce the
wave-vector-dependent spin-glass susceptibility
\begin{equation}
\chi_{\rm SG}\left({\bf k}\right)
= \frac{1}{N}\sum_{i,j} 
\left[\langle S_iS_j\rangle^2\right]_{av} e^{\text{i}{\bf k}
\left({\bf R}_i-{\bf R}_j\right)} .
\end{equation}
The two-point finite-size correlation length $\xi_L$ is then given by
\begin{equation}
\xi_L = \frac{1}{2\sin \left(k_{\rm min}/2\right)} 
\sqrt{\frac{\chi_{\rm SG}({\bf 0})}{\chi_{\rm SG}({\bf k}_{\rm min})} -1},
\label{eq:correlation}
\end{equation}
where ${\bf k}_{\rm min} = (2\pi/L,0,0)$. It scales as
\begin{equation}
\frac{\xi_L}{L}= \tilde{X}[L^{1/\nu}(T-T_c)] ,
\label{eq:xiscale}
\end{equation}
i.e., whenever $T = T_c$ data for different system sizes cross at one
point, up to corrections to scaling.\cite{hasenbusch:08b} 

\section{Numerical Details} 
\label{sec:alg}

To test for universal  behavior a detailed numerical study needs to
be performed where one has to ensure that the data are in thermal
equilibrium.  For this purpose we use a special cluster algorithm
that ensures that spin dimers flip in reasonable simulation times. In
addition, we describe the data analysis used.

\subsection{Algorithm}

The simulations are done using the parallel tempering Monte Carlo
method\cite{hukushima:96} combined with a special cluster flip
algorithm\cite{janzen:08} that ensures ergodic behavior even in the
presence of excessively strong exchange interactions between few spins.

Because the L\'evy distribution has power-law decaying tails, for
certain values of the parameter $\alpha$ the exchange interactions
$J_{ij}$ can be very large.  If two spins have a strong interaction
they will be virtually ``frozen'' under single-spin-flip
dynamics.  To avoid extremely long equilibration times, at the
beginning of each simulation different sets of clusters $C_n$ are
generated.\cite{janzen:08} The generation of the sets $C_n$ is done
the following way:
\begin{enumerate} 

\item{Set $J^0_{\rm min}=T_{\rm max}/4$, where $T_{\rm max}$ is the
maximal temperature from the simulated temperature set.}

\item{The clusters in set $C_n$ consist of spins connected by bonds
that satisfy $\vert J_{ij}\vert > J^n_{\rm min}$.}

\item{The cluster set $C_n$ is stored if $C_n \neq C_{n-1}$ (or
if $n=0$).}

\item{$J^n_{\rm min}$ is iteratively incremented by one ($J^{n+1}_{\rm
min}=J^{n}_{\rm min} +1$). The procedure is repeated initiating from
step two until $C_{n}$ consists only of clusters of size $2$. During
the procedure all clusters are stored.}

\end{enumerate}
One Monte Carlo sweep consists of the following procedure: Each
spin of the system is picked once. After having picked the spin,
a single-spin flip is performed with probability $p=0.75$ (empirically
we find that for $p \sim 0.75$ equilibration is fastest), otherwise
a cluster move is done. In particular:
\begin{itemize}

\item[$\Box$]{{\it The single spin flip} is done with the Metropolis
probability $\min\{1,\exp(-\Delta E/T)\}$, where $\Delta E$ is
the energy difference between the current configuration and the
configuration with the spin flipped.}

\item[$\Box$]{{\it The cluster flip} algorithm works as follows: One
cluster from all sets is randomly (uniformly) picked and flipped with
the Metropolis probability $\min\{1,\exp(-\Delta E/T)\}$, where $\Delta
E$ is the difference between the energy of the actual configuration
and the configuration with the cluster flipped. The cluster flip is
independent of the orientation of the spins in the cluster, i.e.,
the clusters contain only spin indices, such that each spin in the
cluster can change the direction by other update steps.}

\end{itemize}
Note that typical cluster sizes range from $2$ --$20$ spins. 

Because the equilibration test for Gaussian
disorder\cite{katzgraber:01} does not work when the disorder is L\'evy
distributed, the equilibration is monitored by logarithmic binning.
All measured observables (and their higher moments) are recorded as
a function of simulation time. Once the last four bins agree within
error bars the system is deemed to be in thermal equilibrium. If this
test is not passed, the simulation time is increased by a factor of
2 until this is the case. Simulation parameters are summarized in
Table \ref{table:01}.

\begin{table}
\caption{
Parameters of the simulations for different $\alpha$ values. $N_{\rm
sa}$ is the number of samples, $N_{\rm sw}$ is the total number of
Monte Carlo sweeps used for equilibration (the same amount is used
for measurement), $T_{\rm min}$ is the lowest temperature simulated,
$T_{\rm max}$ is the highest temperature simulated, and $N_T$ is the
number of temperatures used in the parallel tempering method for each
system size $L$.
\label{table:01}}
{\footnotesize
\begin{tabular*}{\columnwidth}{@{\extracolsep{\fill}} c r r r r r r}
\hline
\hline
$\alpha$ & $L$ & $N_{\rm sa}$ & $N_{\rm sw}$ & $T_{\rm min}$ & $T_{\rm max}$ & $N_{T}$
\\ 
\hline
$1.00$	& $4$  & $6000$	& $65536$	& $1.112$ & $2.000$ & $12$ \\
$1.00$	& $6$  & $4830$	& $252144$	& $1.112$ & $2.000$ & $12$ \\
$1.00$	& $8$  & $3737$	& $1048576$	& $1.112$ & $2.000$ & $12$ \\
$1.00$	& $10$ & $3400$	& $4194304$	& $1.112$ & $2.000$ & $12$ \\
$1.00$	& $12$ & $3995$	& $16777216$	& $1.112$ & $2.000$ & $12$ \\
$1.00$  & $14$ & $1118$ & $33554432$    & $1.305$ & $1.896$ &  $8$ \\[1mm]

$1.25$	& $4$  & $5600$	& $65536$	& $0.898$ & $1.704$ & $13$ \\
$1.25$	& $6$  & $5082$	& $262144$	& $0.898$ & $1.704$ & $13$ \\
$1.25$	& $8$  & $4165$	& $1048576$	& $0.898$ & $1.704$ & $13$ \\
$1.25$	& $10$ & $4995$	& $2097152$	& $0.898$ & $1.704$ & $13$ \\
$1.25$	& $12$ & $2998$	& $16777216$	& $0.898$ & $1.704$ & $13$ \\[1mm]

$1.50$	& $4$  & $5040$	& $65536$	& $0.726$ & $1.452$ & $14$ \\
$1.50$	& $6$  & $4958$	& $262144$	& $0.726$ & $1.452$ & $14$ \\
$1.50$	& $8$  & $5083$	& $1048576$	& $0.726$ & $1.452$ & $14$ \\
$1.50$	& $10$ & $3014$	& $2097152$	& $0.726$ & $1.452$ & $14$ \\
$1.50$	& $12$ & $3006$	& $16777216$	& $0.726$ & $1.452$ & $14$ \\[1mm]

$1.75$	& $4$  & $5040$	& $65536$	& $0.618$ & $1.305$ & $15$ \\
$1.75$	& $6$  & $5016$	& $262144$	& $0.618$ & $1.305$ & $15$ \\
$1.75$	& $8$  & $4592$	& $1048576$	& $0.618$ & $1.305$ & $15$ \\
$1.75$	& $10$ & $4794$	& $2097152$	& $0.618$ & $1.305$ & $15$ \\
$1.75$	& $12$ & $3999$	& $16777216$	& $0.618$ & $1.305$ & $15$ \\

\hline
\hline
\end{tabular*}
}
\end{table}

\subsection{Finite-size scaling analysis} 

\begin{figure}[!tbh]
\includegraphics[width=\reduce\columnwidth]{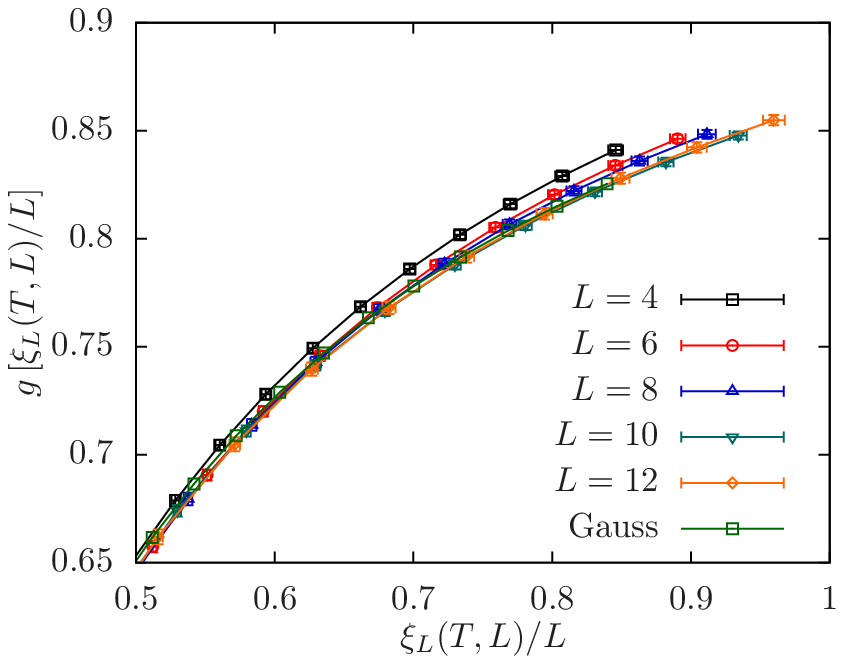}
\caption{(Color online)
Binder ratio $g$ as a function of the finite-size correlation
length $\xi_L/L$ for several system sizes and for a L\'evy parameter
$\alpha=1.25$. Strong corrections to scaling are visible. The data
for the largest system sizes simulated agree with the Gaussian case 
($\alpha = 2$, $L = 24$, from Ref.~\onlinecite{katzgraber:06}).
}
\label{fig:fss_correction}
\end{figure}

To gain insights on the strength of the corrections to scaling, we can
compare two dimensionless quantities,\cite{katzgraber:06,joerg:06} the
correlation length $\xi_L/L$ and the Binder parameter $g$. By plotting
$g[\xi_L(T,L)/L]$ there are no nonuniversal metric factors. Therefore,
data for all system sizes simulated and a given parameter $\alpha$
should all collapse onto a universal function if there are no
corrections to scaling. Data for $\alpha = 1.25$ are shown in Figure
\ref{fig:fss_correction} and illustrate that corrections are large
for $L \lesssim 8$. 

Furthermore, if two different models share the same critical exponent
$\nu$, because no nonuniversal factors when plotting $g[\xi_L(T,L)/L]$
are present, all data should collapse onto a universal curve.  In
Fig.~\ref{fig:fss_correction} we also show data for Gaussian disorder
($\alpha = 2$) for a large system size ($L = 24$).\cite{katzgraber:06}
Data for $\alpha = 1.25$ and $L \gtrsim 8$ agree with the Gaussian
case, thus illustrating that for a conventional scaling analysis only
the largest system sizes should be included.

We have attempted different scaling
approaches,\cite{joerg:06,katzgraber:06} as well as the inclusion
of corrections to scaling.\cite{hasenbusch:08b} However, large system
sizes are difficult to simulate for L\'evy spin glasses and therefore
we use the  extended scaling technique\cite{campbell:06} that allows
us to include smaller system sizes in the scaling analysis.

\begin{figure*}[!tbh]

\includegraphics[width=0.30\textwidth]{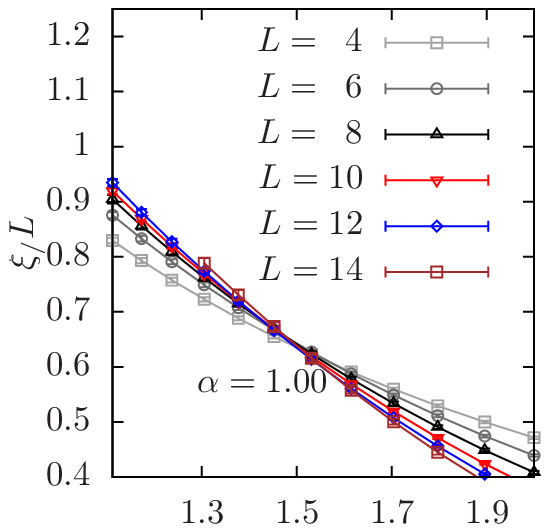}
\includegraphics[width=0.30\textwidth]{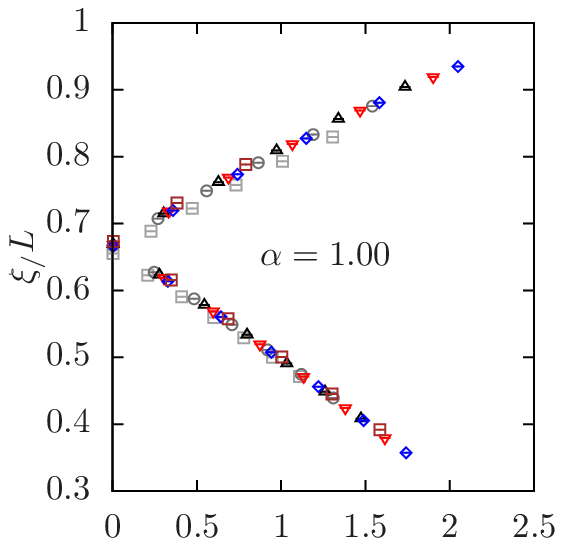}
\includegraphics[width=0.30\textwidth]{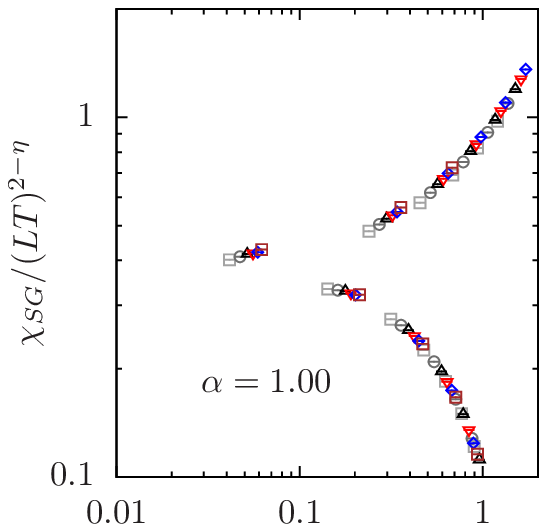}

\vspace*{0.3em}

\includegraphics[width=0.30\textwidth]{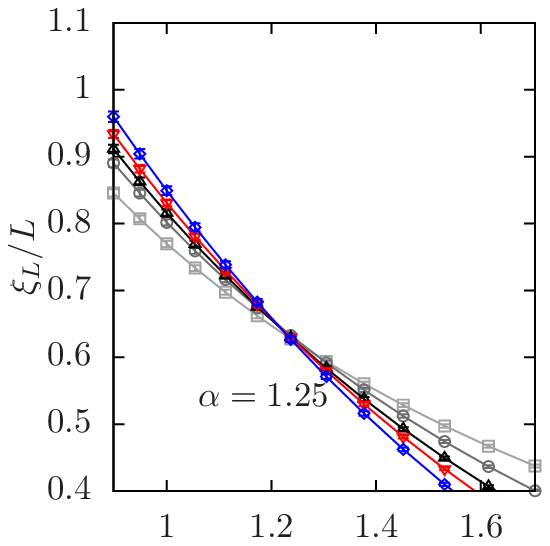}
\includegraphics[width=0.30\textwidth]{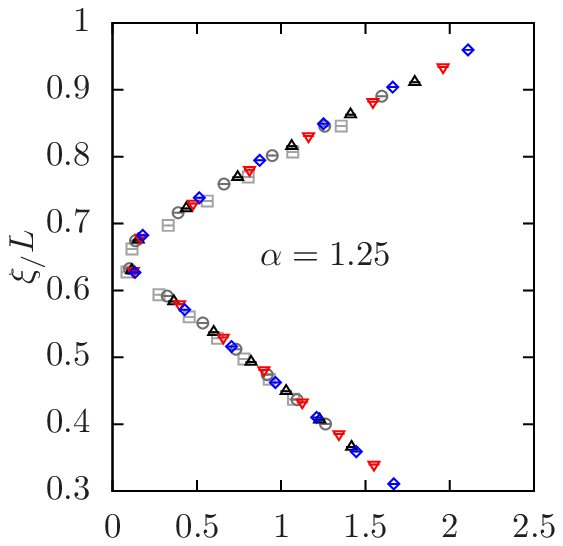}
\includegraphics[width=0.30\textwidth]{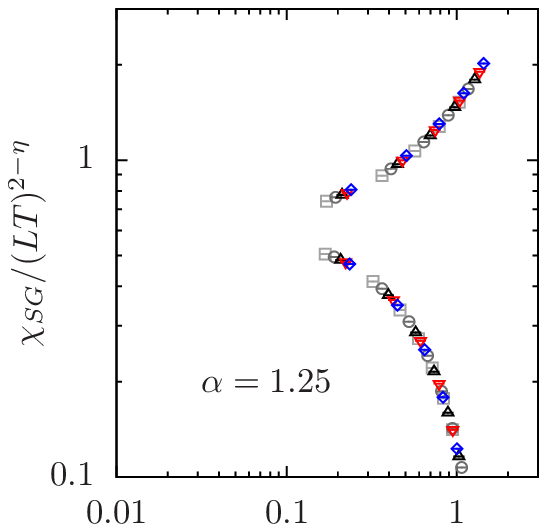}

\vspace*{0.3em}

\includegraphics[width=0.30\textwidth]{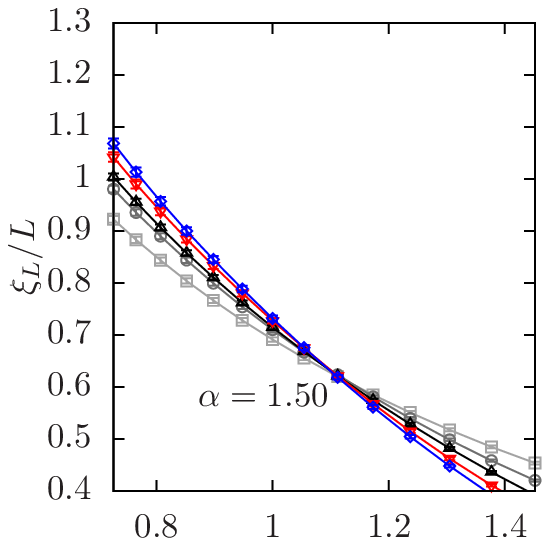}
\includegraphics[width=0.30\textwidth]{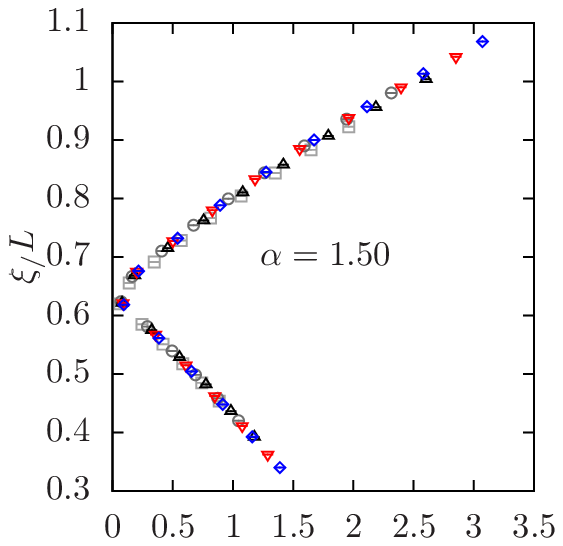}
\includegraphics[width=0.30\textwidth]{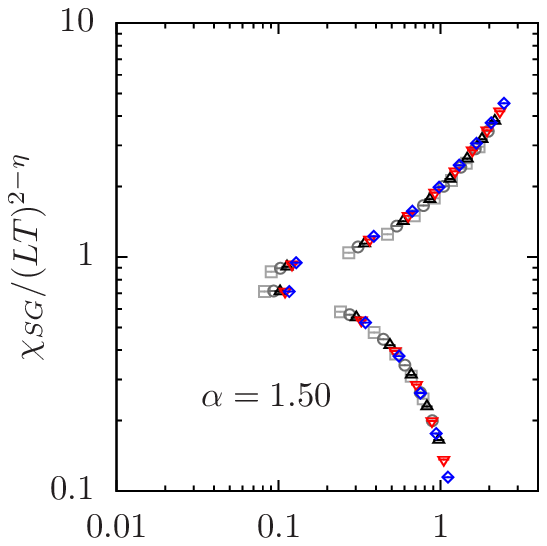}

\vspace*{0.3em}

\includegraphics[width=0.30\textwidth]{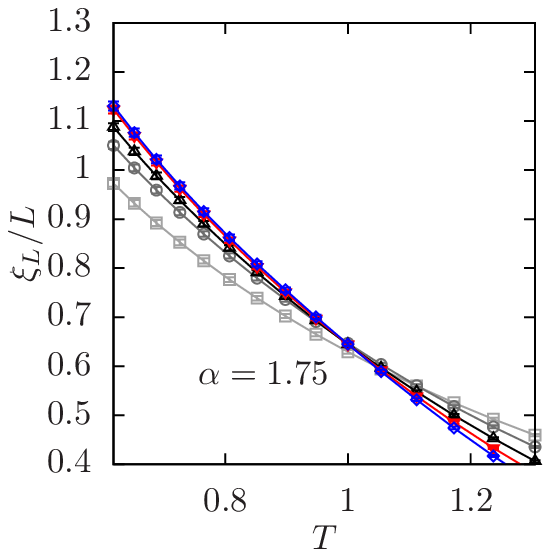}
\includegraphics[width=0.30\textwidth]{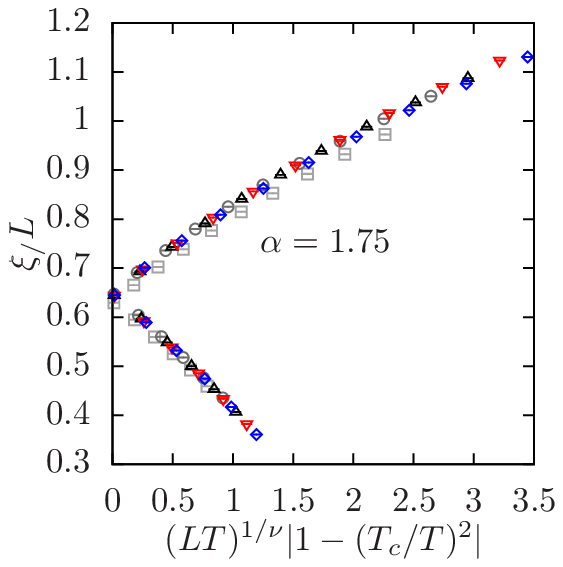}
\includegraphics[width=0.30\textwidth]{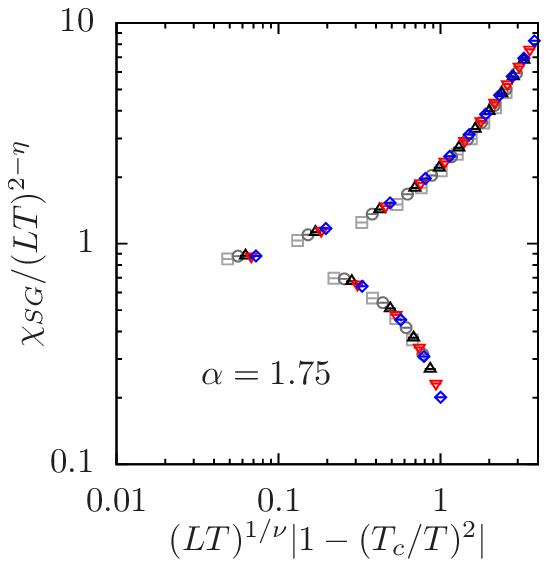}

\vspace*{-.2cm}

\caption{(Color online)
Left: Two-point finite-size correlation length $\xi_L/L$ as a function
of the temperature $T$ for different L\'evy parameters $\alpha$. The
data cross, thus signaling the presence of a transition.  Center:
Extended scaling of the two-point finite-size correlation length
for different $\alpha$. Right: Extended scaling of the spin-glass
susceptibility for different $\alpha$. The data scale very well and
the critical exponents extracted from the scaling agree within error
bars, thus suggesting that all systems share the same universality
class. See Table \ref{table:02} for the optimal scaling parameters.
}
\label{fig:scaling}
\end{figure*}

Within the extended scaling framework\cite{campbell:06} the standard
scaling expression for the correlation length, Eq.~(\ref{eq:xiscale}),
is replaced by
\begin{equation}
\frac{\xi_L}{L}\sim 
\tilde{X}[(LT)^{1/\nu}| 1-(T/T_c)^2|] .
\label{eq:xiextended}
\end{equation}
The aforementioned expression is derived by including a resummation
of a high-temperature series expansion and therefore includes effects
of corrections to scaling. Similarly, the scaling relation for the
susceptibility, Eq.~(\ref{eq:susceptibility}), is replaced by
\begin{equation}
\chi_{SG}(L,T) \sim 
(LT)^{2-\eta}
\tilde{C}[(LT)^{1/\nu}| 1-(T/T_c)^2|] .
\label{eq:susceptibilityextended}
\end{equation}
We assume that the scaling function in Eq.~(\ref{eq:xiextended}) can be
approximated by a third-order polynomial for temperatures larger than
$T_c$, i.e., $\tilde{X}(x) =a+bx+cx^2+dx^3$ where $x=(LT)^{1/\nu}|
1-(T/T_c)^2|$, $T>T_c$  and perform a fit to the six parameters $a$,
$b$, $c$, $d$, $T_c$ and $\nu$. A similar approach is used for the
spin-glass susceptibility [Eq.~(\ref{eq:susceptibilityextended})],
for which there is a seventh parameter, the critical exponent $\eta$.
Note that the extended scaling scheme only works for temperatures $T
> T_c$.  Therefore, we first perform a rough estimate of $T_c$ using
conventional scaling methods.  The nonlinear fit is performed with
the statistics package R,\cite{R} including system sizes $L\geqslant
6$. Error bars are determined using a bootstrap analysis.

To compute the error bars we apply the following procedure: For each
system size $L$ and $N_{\rm sa}$ disorder realizations, a randomly
selected bootstrap sample of $N_{\rm sa}$ disorder realizations is
generated. With this random sample, an estimate of the different
observables is computed for each temperature. We repeat this procedure
$N_{\rm boot} = 500$ times for each lattice size and then assemble
$N_{\rm boot}$ complete data sets (each having results for every
size) by combining the $i$-th bootstrap sample for each size for
$i=1,\ldots,N_{\rm boot}$. The finite-size scaling fit described
above is then carried out on each of these $N_{\rm boot}$ sets, thus
obtaining $N_{\rm boot}$ estimates of the fit parameters. Because the
bootstrap sampling is done with respect to the disorder realizations
which are statistically independent, we can use a conventional
bootstrap analysis to estimate statistical error bars on the fit
parameters. These are comparable to the standard deviation among the
$N_{\rm boot}$ bootstrap estimates.\cite{katzgraber:06}

From the aforementioned finite-size scaling we can also
extract the critical temperature $T_c(\alpha)$ to compare to
analytical predictions. We use the critical temperature estimated
using the extended scaling method for the correlation length
[Eq.~(\ref{eq:xiextended})] because corrections to scaling are
smaller than for the spin-glass susceptibility. Furthermore, the
bootstrap analysis requires one parameter less leading to smaller
statistical errors.

\section{Results} 
\label{sec:results}

\begin{figure}[!tbh]
\includegraphics[width=\reduce\columnwidth]{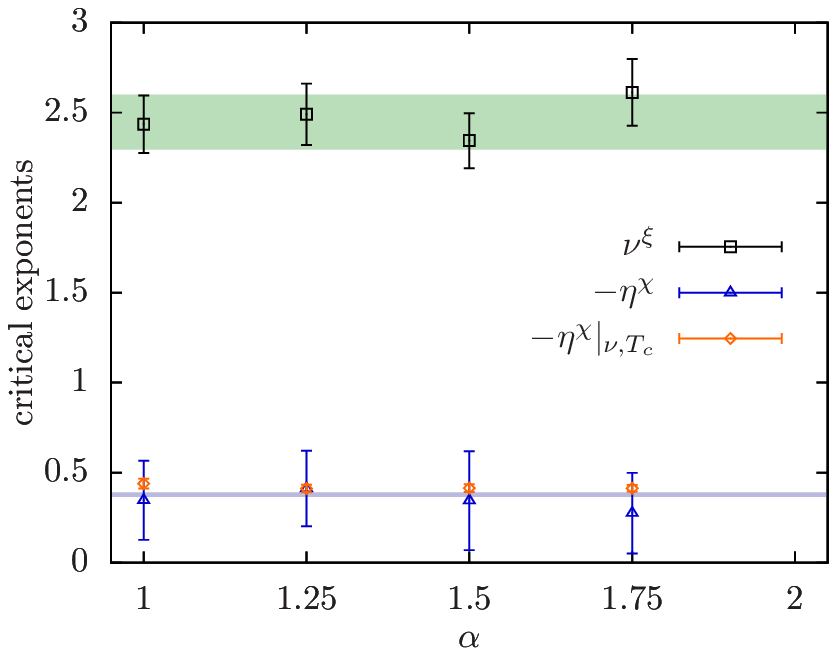}
\caption{(Color online)
Critical exponents $\eta$ and $\nu$ as a function of the L\'evy
parameter $\alpha$. The shaded areas correspond to the estimates for
Bimodal disorder from Ref.~\onlinecite{hasenbusch:08b}.  The estimate
for the critical exponent $\nu^\xi$ comes from an extended finite-size
scaling analysis of the two-point correlation length. The estimates
for the critical exponent $\eta$ are from two independent analyses
of the spin-glass susceptibility. $\eta^\chi$ is computed from
an extended finite-size scaling analysis where $\eta$, $\nu$, and
$T_c$ are parameters, whereas $\eta^\chi|_{\nu,T_c}$ is computed by
fixing $\nu = \nu^\xi$ and $T_c = T_c^\xi$ from the analysis of the
two-point correlation length.  For all values of $\alpha$ studied,
the exponents $\nu^\xi$ and $\eta^\chi$ are in good agreement with the
best-known estimates for the bimodal case.  However, the estimate for
$\eta^\chi\vert_{\nu,T_c}$ consistently lies above the best estimate
for $\eta$ possibly due to strong corrections to scaling that we
cannot account for, as well as systematic errors from the determination
of $\nu^\xi$.\cite{katzgraber:06}
}
\label{fig:critical_exponents}
\end{figure}

Corrections to scaling for small systems of L\'evy spin glasses with
$1\leqslant\alpha<2$ are large (see Fig.~\ref{fig:fss_correction}).
We attempt to scale the data using the extended scaling scheme,
as shown in Fig.~\ref{fig:scaling} (center and right columns). The
left column shows the finite-size correlation length for different
values of the parameter $\alpha$. In all cases the data cross at a
transition temperature that decreases with increasing $\alpha$. In
the center panels of Fig.~\ref{fig:scaling} we show an extended
scaling of the two-point finite size correlation length according
to Eq.~(\ref{eq:xiextended}) with the critical exponents $\nu$
and $T_c$ as parameters. The right column of Fig.~\ref{fig:scaling}
shows an extended scaling of the spin-glass susceptibility according
to Eq.~(\ref{eq:susceptibilityextended}) with $\eta$, $\nu$ and
$T_c$ as free parameters.  The scaling of the data works well and,
in particular, the estimated critical exponents agree with the bimodal
values.\cite{hasenbusch:08b} Our best estimates are summarized in Table
\ref{table:02}. Furthermore, in Fig.~\ref{fig:critical_exponents} we
compare our estimates for $\eta$ and $\nu$ to the bimodal estimates
[$\eta = -0.375(10)$ and $\nu = 2.45(15)$].\cite{hasenbusch:08b}
The data therefore suggest that all studied L\'evy spin glasses share
the same universality class.

\begin{table}
\caption{
Summary of estimates of the critical parameters. $T_c^\xi$ and
$\nu^\xi$ are the critical parameters estimated from an extended
scaling analysis of the two-point correlation length, whereas
$\eta^\chi$ has been computed from a finite-size scaling analysis of
the spin-glass susceptibility with $T_c$, $\nu$ and $\eta$ as free
parameters.  $\eta^\chi|_{T_c,\nu}$ is computed from a finite-size
scaling analysis of the susceptibility with $T_c = T_c^\xi$ and $\nu =
\nu^\xi$ fixed and only $\eta$ as a parameter.
\label{table:02}}
{\footnotesize
\begin{tabular*}{\columnwidth}{@{\extracolsep{\fill}} c l l l l l}
\hline
\hline
$\alpha$ & $T_c^\xi$   & $\nu^\xi$  & $\eta^\chi$   & $\eta^\chi|_{T_c,\nu}$ \\
\hline
$1.00$	 & $1.467(31)$ & $2.42(17)$ & $-0.346(220)$ & $-0.438(26)$  & \\
$1.25$	 & $1.209(28)$ & $2.49(17)$ & $-0.411(209)$ & $-0.412(20)$  & \\
$1.50$	 & $1.094(21)$ & $2.34(15)$ & $-0.344(274)$ & $-0.414(20)$ & \\
$1.75$	 & $0.996(20)$ & $2.61(19)$ & $-0.274(224)$ & $-0.413(17)$ & \\
\hline
\hline
\end{tabular*}
}
\end{table}

\begin{figure}[!tbh]
\vspace*{-0.3cm}
\includegraphics[width=\reduce\columnwidth]{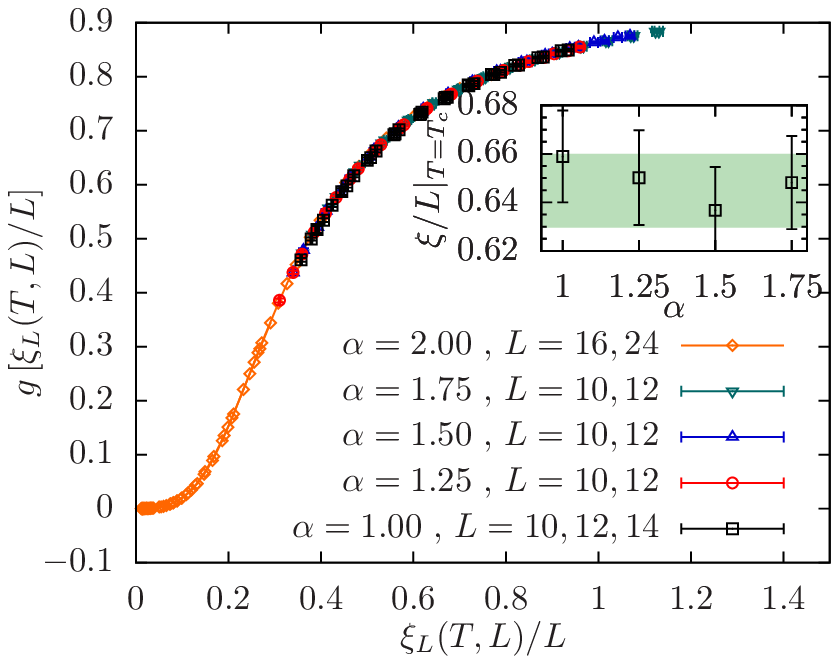}
\caption{(Color online)
Binder ratio $g$ as a function of the two-point finite-size correlation
length $\xi_L/L$ for $\alpha = 1.00$, $1.25$, $1.50$ and $1.75$ for
system sizes $L=10$ and $12$ ($14$, for $\alpha = 1.00$), as well as
$16$ and $24$ for the Gaussian ($\alpha=2.0$) case.\cite{katzgraber:06}
The line is a guide to the eye.  All data collapse onto a
universal curve, thus providing further evidence for universality.
The inset shows $\xi_L/L\left(T=T_c\right)$---also a
universal quantity---as a function of $\alpha$. For all values of
$\alpha$ studied the data agree within error bars. The horizontal
shaded area corresponds to the best estimate for bimodal disorder
$\xi_L/L\left(T=T_c\right) = 0.645(15)$.\cite{hasenbusch:08b}
}
\label{fig:universality}
\end{figure}

To further strengthen our results for the finite-size correlation
length, in Fig.~\ref{fig:universality} we show $g[\xi_L(L,T)/L]$ for
the largest system size studied and different $\alpha$, as well as
data for Gaussian disorder.\cite{katzgraber:06} The data collapse
cleanly onto a universal curve without any scaling parameters
providing further evidence for universal behavior. The inset of
Fig.~\ref{fig:universality} shows $\xi_L/L\left(T=T_c\right)$
for different values of the exponent $\alpha$. For all cases the
data agree within error bars with the best estimate for bimodal
disorder\cite{hasenbusch:08b} hence strengthening our claim for
universal behavior.

\begin{figure}[!tbh]
\includegraphics[width=\reduce\columnwidth]{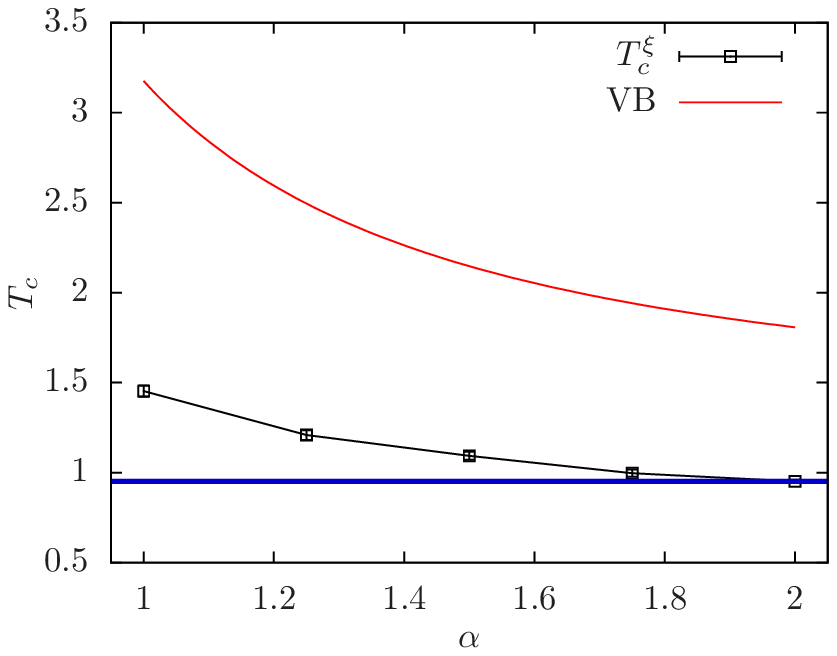}
\caption{(Color online)
Critical temperature $T_c$ computed from a finite-size scaling
of the two-point correlation length, Eq.~(\ref{eq:correlation}),
using the extended scaling technique, Eq.~(\ref{eq:xiextended}).
The continuous curve (labeled with VB) is the critical temperature
$T_c$ for a mean-field spin glass on a diluted graph with fixed
connectivity $k+1=6$. The horizontal thick line represents the critical
temperature $T_c$ for the Gaussian spin glass.\cite{katzgraber:06}
}
\label{fig:t_c}
\end{figure}

Finally, we show in Fig.~\ref{fig:t_c} estimates for the
critical temperature $T_c$ as a function of the exponent
$\alpha$.\cite{janzen:08,neri:10} The horizontal blue line represents
the Gaussian limit.\cite{katzgraber:06,hasenbusch:08b} The red curve
represents $T_c(\alpha)$ for a mean-field spin-glass model on a
diluted graph with fixed connectivity \mbox{$k+1=6$}. The critical
temperature is determined from the following equation
\begin{equation}
\label{eq:mean_field_Tc}
1 = k \int dJ {\mathcal P}(J)\tanh^2(J/T_c), 
\end{equation}
where ${\mathcal P}(J)$ is given by
Eq.~(\ref{eq:distribution}).\cite{thouless:86} There is qualitative
agreement in the trend of the data.  At first sight, there is a
disagreement to the behavior obtained for the infinite-range model
studied in Refs.~\onlinecite{cizeau:93} and \onlinecite{janzen:08},
because the dependence on $\alpha$ is reversed in that case.
However, the large connectivity limit of Eq.~(\ref{eq:mean_field_Tc})
amounts to the expression for the critical temperature stated in
Refs.~\onlinecite{cizeau:93} and \onlinecite{janzen:08}.  For the
infinite-range model an $\alpha$-dependent rescaling of the couplings
(\mbox{$J_{ij} \to J_{ij} N^{-1/\alpha}$}) is necessary to obtain a
non-trivial thermodynamic limit which changes the energy scale in an
$\alpha$-dependent way.

\section{Conclusions} 

We have studied the critical behavior of a three-dimensional Ising
spin glass with L\'evy-distributed interactions to test universality.
An extended scaling analysis of the correlation length and spin-glass
susceptibility suggests that for all values of $\alpha$ the L\'evy spin
glass obeys universality.  Previous claims that universality might
be destroyed when $\alpha \to 1$ possibly stem from the fact that
the simulations did not take into account the effects of the strong
interactions between some spins, i.e., rendering the simulations
nonergodic.  Further support for universal behavior is given by the
plot of $g$ against $\xi_L/L$, Fig.~\ref{fig:universality}, where data
for all the models studied collapse onto a single universal curve.
We do find strong corrections to scaling and therefore studies with
larger system sizes and a clear understanding of scaling corrections
would be desirable.  However, we find no clear evidence for the lack
of universality.

\begin{acknowledgments} 

We thank A.~K.~Hartmann for numerous discussions. H.G.K.~acknowledges
support from the SNF (Grant No.~PP002-114713). The authors acknowledge
ETH Zurich for CPU time on the Brutus cluster.

\end{acknowledgments}

\bibliography{refs}
 
\end{document}